\documentclass{vgtc}                          

\graphicspath{{figures/}{pictures/}{images/}{./}} 

\usepackage{times}                     

\usepackage{tabu}                      
\usepackage{booktabs}                  
\usepackage{lipsum}                    
\usepackage{mwe}                       

\usepackage{mathptmx}                  

\usepackage{color}
\usepackage[percent]{overpic}

\let\oldput\put
\def\put(#1,#2)#3{%
  \oldput(#1,#2){\sffamily #3}%
}
\usepackage[all]{nowidow}

\onlineid{1042}

\vgtccategory{Research}

\vgtcinsertpkg

\title{Bimanual XR Specification of Relative and Absolute Assembly Hierarchies for Teleoperation}

\author{Ben Yang* %
\and Xichen He* %
\and Charlie Zou*  %
\and Jen-Shuo Liu*  %
\and Barbara Tversky*  %
\and Steven Feiner\thanks{\{by2297, xh2623, jz3331, jl5004, bt2158, feiner\}@columbia.edu}
}
\affiliation{\scriptsize Columbia University}

\teaser{
  \centering
      {\setlength{\fboxsep}{1pt}
    \framebox{\begin{overpic}[width=.235\linewidth, trim=0 0cm 0 0cm, clip]{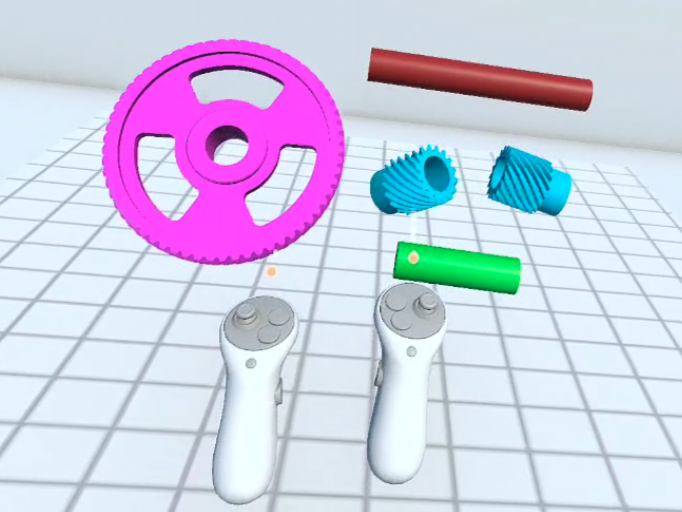}
    \put(1,2){\small{\color{black} \textbf{(a)}}}
    \end{overpic}}}
    {\setlength{\fboxsep}{1pt}
    \framebox{\begin{overpic}[width=.235\linewidth, trim=0 0cm 0 0cm, clip]{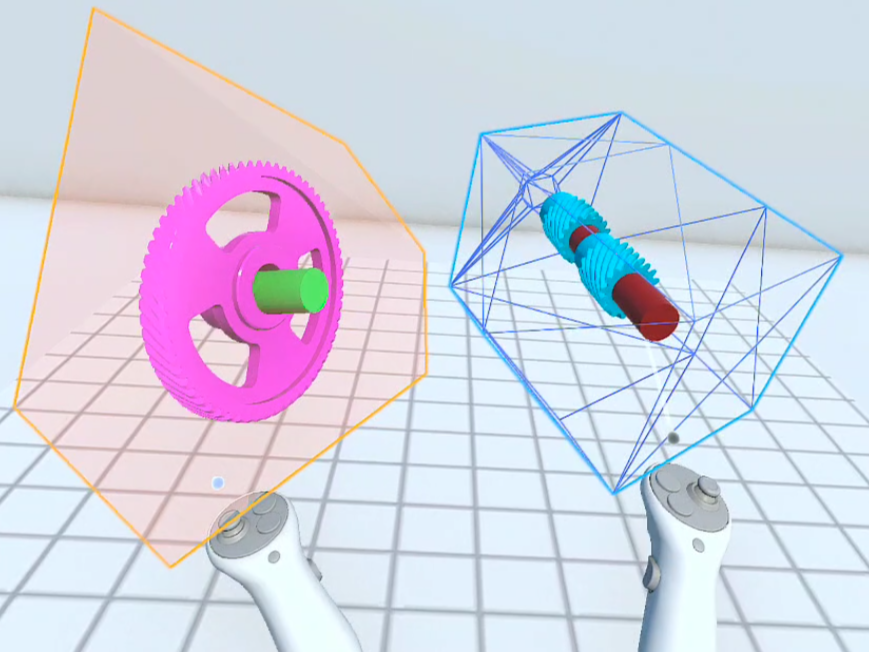}
    \put(1,2){\small{\color{black} \textbf{(b)}}}
    \end{overpic}}}
    {\setlength{\fboxsep}{1pt}
    \framebox{\begin{overpic}[width=.235\linewidth, trim=0 0cm 0 0cm, clip]{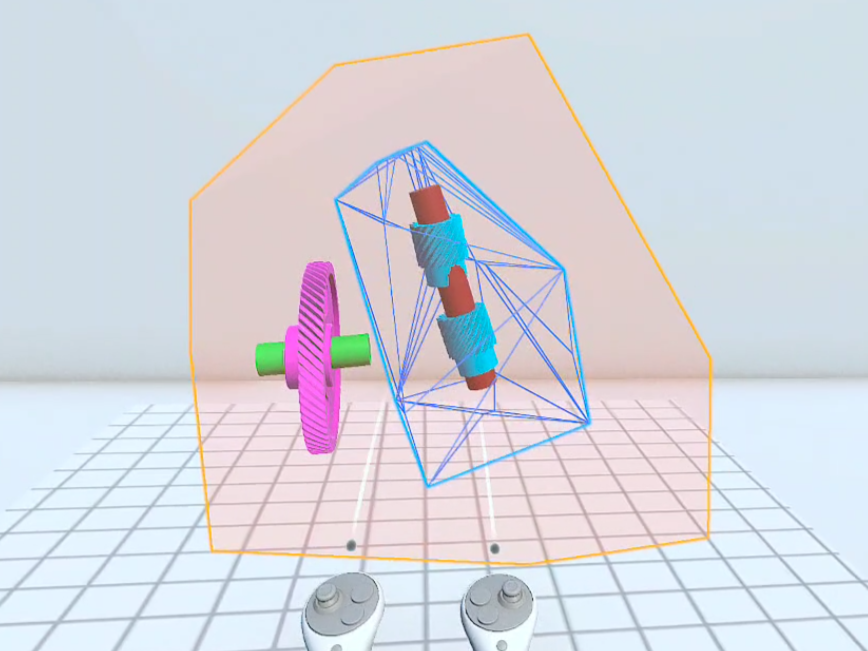}
    \put(1,2){\small{\color{black} \textbf{(c)}}}
    \end{overpic}}}
    {\setlength{\fboxsep}{1pt}
    \framebox{\begin{overpic}[width=.235\linewidth, trim=0 0cm 0 0cm, clip]{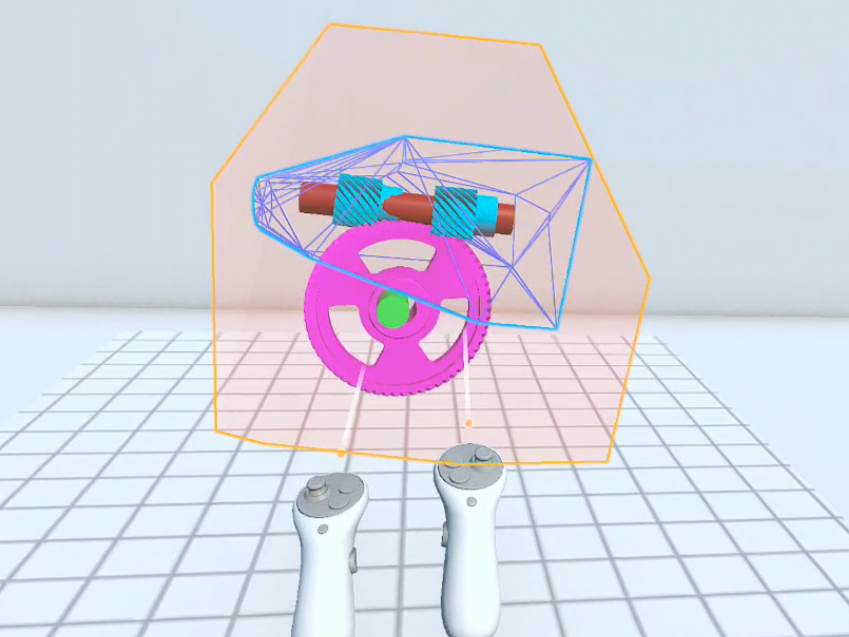}
    \put(1,2){\small{\color{black} \textbf{(d)}}}
    \end{overpic}}}
  \caption{Bimanual authoring of assembly hierarchies.
  (a) Virtual replicas of gear train components in the XR workspace: two shafts and three gears.
  (b) User grabs one object in each hand to create a constraint group (two of which are shown). A relative group (orange semi-transparent hull) is created around the input shaft and one gear; an absolute group (blue wireframe hull) is created around the output shaft and two gears. The relative group can be placed by the robot; the absolute group maintains a fixed pose.
  (c) The user grips both hulls to nest the absolute group inside the relative group, creating a parent--child hierarchy. The wireframe child hull appears inside the semi-transparent parent hull.
  (d) The user specifies the absolute child group's 6DoF pose within the relative parent group whose 6DoF pose can be determined by the robot.
  Even if the relative parent group is moved, the absolute child group maintains its 6DoF pose in relation to the parent group.
  }
  \label{fig:teaser}
}

\abstract{
We present a bimanual XR interaction approach for specifying remote assembly tasks as hierarchies of relative and absolute object constraints that specify high-level teleoperation goals for robots. Grabbing one object in each hand creates a constraint group (visualized as a hull) and groups can be nested into hierarchies. Each group can be relative (with a robot-specifiable 6DoF pose) or absolute (with an author-specified fixed 6DoF pose) in relation to its parent. A relative group specifies a subassembly that can be constructed at a location chosen by the robot software for efficiency rather than mandated by the user.

} 

\keywords{Extended reality, bimanual grouping, constraint specification, robot teleoperation.}

\begin{document}


\firstsection{Introduction}

\maketitle
Extended reality (XR) enables intuitive specification of robot assembly tasks through virtual replicas~\cite{aoyama_asynchronously_2024} of physical objects. In this paradigm, a user specifies assembly constraints (spatial relationships between objects or the world) in XR for a robot to execute at the physical site. Prior work introduced \textit{relative} constraint specification, where object poses are specified only with respect to each other rather than absolute world coordinates~\cite{yang_relative_2025}. When only inter-object relationships matter, robots can build subassemblies anywhere in the workspace and specify them relative to other components only when needed. When exact 6DoF poses matter, absolute constraints can fix objects to specific poses.

Much prior work on XR-based robot task specification~\cite{aoyama_asynchronously_2024, arboleda_assisting_2021, cai_hierarchical_2024, cheng_open-television_2024, yang_relative_2025} has two limitations. First, sequential object selection requires users to select objects one at a time to group them, rather than simultaneously grabbing objects to indicate grouping. Second, single-level grouping limits the specification of assemblies where subassemblies must be built before combining them (e.g., attaching gears, mounting shaft assemblies in a housing). 

We address these limitations by using bimanual interaction and supporting hierarchical constraint groups. Our key observation is that grabbing one object in each hand can represent the intent to join them, analogous to physically holding parts before joining. This action indicates that these objects should be assembled together; additional objects (e.g., a screw to fasten two plates) can be added by grabbing while holding a grouped object. Groups are visualized as
convex hulls around padded object geometry. Groups can be nested into hierarchies by grabbing their hulls, with parent hulls visually encompassing children to represent nested subassemblies. This hierarchical structure enables a robot planner to stipulate assembly sequences: subassemblies can be built independently before integration. The resulting constraints are exported for robot execution through  simulation and motion planning.
Thus, we make the following contributions:
\begin{itemize}
    \vspace{-6pt}
    \setlength\itemsep{-2pt}
    \item A bimanual interaction technique where grabbing one object in each hand creates constraint groups, with additional objects added by grabbing while holding a grouped object.
    \item A hierarchical constraint system with nested hull visualizations and hull-visibility controls for intuitive operation.
    \item A pipeline translating hierarchical relative constraints to simulated robot execution using inter-group and within-group optimizations.
\end{itemize}

\section{Related Work}
\label{sec:related}

\textbf{Remote Task Guidance.}
XR systems for remote guidance have combined video and 3D reconstruction for shared spatial understanding~\cite{teo_mixed_2019}, while AR visual cues can assist manipulation in robot teleoperation~\cite{arboleda_assisting_2021}. For robot assembly, virtual replicas enable goal specification~\cite{aoyama_asynchronously_2024}, and relative constraint specification uses hull visualizations~\cite{yang_relative_2025}. We extend this with bimanual interaction for group creation and hierarchical constraint nesting.

\textbf{Bimanual XR Interaction.}
Voodoo Dolls~\cite{pierce_voodoo_1999} uses bimanual manipulation of temporary proxy objects for precise pose specification, with the non-preferred hand providing a reference frame for the preferred hand. Recent work uses bimanual gestures for 3D content authoring with gesture-based grammars~\cite{lee_wiresketch_2022, lee_bimanual_2024}. We also use bimanual input for the creation of constraint groups and their arrangement into hierarchies.

\textbf{Robot Task Specification using XR.}
Previous work ranges from low-level direct teleoperation~\cite{cheng_open-television_2024} to hierarchical approaches in which high-level task intent informs low-level action generation~\cite{cai_hierarchical_2024, ma_hdp_2024}. We enable users to specify the spatial relationships that \textit{must} hold through XR authoring; our robot planner determines the remaining ones, and plans assembly order with collision-free paths.

\section{System Description}
\label{sec:system}

\subsection{Constraint Authoring}

A user can specify assembly relationships through bimanual interaction, as shown in \cref{fig:teaser}. The controller trigger button is used to select an object, while the grip button is used to select a hull, allowing a controller selection ray to penetrate hull colliders for nested selection. The user creates a constraint group by simultaneously selecting one object from those available (\cref{fig:teaser}a) in each hand using the trigger buttons; a hull appears around grouped objects as visual feedback. \cref{fig:teaser}(b) shows a relative group and an absolute group, as described below. A group containing only a single object can be created by selecting that object with both hands.

Groups can be nested hierarchically (\cref{fig:teaser}c--d) by gripping two hulls. The hull that is gripped first becomes the parent of the hull that is gripped second, enabling compositional specification of complex assemblies, with no limit on nesting depth. The parent hull visually encompasses its child hulls. Gripping a hull and grabbing an ungrouped object with the trigger button adds that object to the hull's group. Gripping a hull with one controller and grabbing an object that belongs to a different group with the trigger button of the other controller creates a new parent hull containing both groups as children. To reduce clutter, the user can apply hull-visibility controls so that only outermost hulls are shown by default, revealing child hulls when the controller moves inside a hull. 
The poses of the children of a group remain editable until the user exports the group to the robot pipeline for execution. The user can delete a hull by pressing a designated button while gripping the hull; the hull's children then become children of the deleted hull's parent.

A group can be toggled between two modes: \textit{relative} (with a robot-specificable 6DoF pose) and \textit{absolute} (with a user-specified fixed 6DoF pose) in relation to its parent by pressing a button while gripping the hull. A \textit{relative} group is surrounded by a semi-transparent hull; an \textit{absolute} group is surrounded by a wireframe hull. Relative versus absolute determines the group's relationship to its parent. Every object has an absolute 6DoF pose in relation to its parent.  For example, a \textit{relative} gear-train group inside an \textit{absolute} fixture group can have its 6DoF pose set by the robot, but the 6DOF pose of the fixture group remains locked to its parent.


\subsection{Robot Execution Pipeline}
Our system exports constraint specifications, capturing 6DoF poses, group modes, and hierarchical relationships. \textit{Relative} groups allow the robot planner to optimize 6DoF poses using nonlinear optimization\footnote{\url{https://drake.mit.edu}} subject to collision avoidance, gravity, and reachability constraints; \textit{absolute} groups enforce exact 6DoF poses specified during XR authoring. We use Traveling Salesperson Problem-based optimization to order group visitation to minimize robot travel distance between group centroids, and within-group object sequencing to minimize total joint-space travel.


MuJoCo\footnote{\url{https://mujoco.org}} simulates contact dynamics and gravity, with a physics settling phase that resolves interpenetrations from XR pose specification. MoveIt\footnote{\url{https://moveit.ros.org}} plans collision-free paths. The same code transfers to physical hardware by replacing the simulation with the robot driver, allowing validation before deployment.

\subsection{Implementation}
Our system is implemented in Unity 6.2 for Quest 3, connected with Quest Link to a Windows 11 computer with an Intel Core i9-9900K and an NVIDIA GeForce RTX 4090. Hulls are computed using a GPU-parallel custom Quickhull\footnote{\url{https://qhull.org/}} variant with spatial hash clustering for O($n$) vertex reduction and parallel apex finding that searches neighboring grid cells, achieving 3--10x speedup over CPU implementations. The robot execution pipeline runs on Ubuntu 24.04 with MoveIt and MuJoCo, simulating a Universal Robots UR5 robot arm.

\section{Discussion and Future Work}
\label{sec:observations}
In robot execution, the robot autonomously determines the execution order; the MuJoCo physics settling phase resolves minor interpenetrations from the XR pose specification. A limitation of our current visualization is that hulls can include empty regions not occupied by group members (e.g., the center of four objects arranged in a square); ungrouped objects in these regions may appear visually grouped. 

We presented a bimanual XR technique for authoring robot assembly: simultaneously selecting objects creates constraint groups with relative or absolute modes, and nesting these groups creates hierarchies for assembly ordering. Our future work includes a formal user study comparing the advantages and disadvantages of relative vs. absolute groups.

\acknowledgments{
This research was funded by National Science Foundation Grant CMMI-2037101.}

\bibliographystyle{abbrv-doi}

\bibliography{template}
\end{document}